\documentclass[preprint,12pt]{elsarticle}
\usepackage{amssymb}
\usepackage{epsfig}
\usepackage{amsthm}
\usepackage{amsmath}
\journal{Physica A}
\begin{document}
\begin{frontmatter}

\title{Analytical expression for end-to-end- auto correlation function of a long chain polymer molecule in solution}

\author{Moumita Ganguly \corref{cor1}\fnref{fn1}}
\ead{mouganguly09@gmail.com}
\address{Indian Institute of Technology Mandi, Kamand, Himachal Pradesh - 175005, India.}

\author{Anirudhha Chakraborty\corref{cor2}}
\address{Indian Institute of Technology Mandi, Kamand, Himachal Pradesh - 175005, India.}
\cortext[cor1]{Corresponding author}

\fntext[fn1]{Phone:-+91-1905-267145, Fax: +91-1905-237945}

\begin{abstract} 
\noindent A diffusion-like theory for real time end-to-end distance of a long polymer chain in dilute solution is formulated. We give a detailed analytical expression for the end-to-end distance auto-correlation function of a long chain polymer in solution. The physical problem of dynamics of end-to-end distance can be modeled mathematically with the use of a Smoluchowski-like equation. Using this equation analytical expression for end-to-end distance auto-correlation function is derived. We find that this auto-correlation function varies with several parameters such as length of the polymer (N), bond length (b) and the relaxation time ${\tau_R}$. 
\end{abstract}


\end{frontmatter}

\section{Introduction}

\noindent Understanding the dynamics end-to-end distance of a long chain polymer molecules in solution  has been an interesting research field \cite{Szabo-Physica, M-Physica, M-CPL}. This will help a lot in understanding polymer looping problem both experimentally \cite{Winnik,Haung,Lapidus} and theoretically \cite{Wilemski,Doi,Pastor,Portman,Sokolov,Toan}. In this paper, the dynamics of end-to-end distance of a single long  polymer chain have been modeled following the work of Szabo {\it et. al.,} \cite{Szabo}. In this model dynamics of end-to end distance is mathematically represented by a Smoluchowski-like equation for a single particle under harmonic potential. 

\section{Our Model}
\noindent In this section we use the simplest one dimensional description of a polymer as given by Szabo {\it et.al.,} \cite{Szabo}. Our description consists of a total $2N$ segments of unit length. The chain consists of $2N$ monomers following random walk. Thus each monomer is allowed to take two possible orientations, one along the right and the other among the left direction. So the polymer can have any of the total $2^{2N}$ different conformations. In the following, we define $x$ as the end-to-end distance of the whole polymer, with the value $x=2j$. After a total $N$ steps the polymer segments can be either on the right $N+j$ or on the left $N-j$. This implies that the polymer looping problem can be solved using standard procedure of probability theory. The equilibrium end-to-end distribution $P_{0j}$ of this long polymer is given by
\begin{equation}
P_{0j}=2^{-2N}(^{2N}_{N+j}).
\end{equation}
Now if we consider the unbiased random walk problem, then the probability distribution can be very easily calculated by knowing the differences of rate of fluctuations between the polymers moving towards either right or left. If we imagine the polymer molecule to be immersed in a solvent, the motion of polymer will be determined by a wide variety of intra-molecular and inter-molecular forces between the polymer and the solvent. When we watch the motion of polymer molecule, the motion would appear to be random. Then considering the variation of all right and left monomer segments being independent of each other the net fluctuation is directed by the following rate equation.
\begin{equation}
\label{2}\frac{d}{dt}\left[^p_n\right]=\frac{1}{\tau_R}\left[
\begin{array}{cc}
-1&1\\
1& -1
\end{array}
\right]\left[^p_n\right],
\end{equation}
where the vector $[^p_n]$ represents the activity of right and left segments orientations, $\tau_R$ represent the relaxation time from one configuration to another. Now, the whole event of end-to-end looping of a polymer molecule in solution, can be considered to be a simple random walk model confined in $2^{2N}$-dimensional configuration space. The individual monomer's re-orientation would result in $2N$ ways to reorder a $x=2j$ conformation either to a $x=2j+2$ or $x=2j-2$ conformation and $N-J+1$ ways to reorient a $x=2j+2$ conformation into a $x=2j$ conformation. Then the resulting master equation for the end-to-end distribution $P(j,t)$ in the $(2N+1)$-dimensional space is given by \cite{Szabo}
\begin{equation}
\tau_R\frac{d}{dt}P(j,t)=-2NP(j,t)+(N+j+1)P(j+1,t)+(N-j+1)P(j-1,t).
\end{equation}
\noindent As we see that for long chain polymer molecule ($N$ large), what we try to search for and measure experimentally is a distribution, {\it{i.e.,}} the probability for finding the end-to-end distance of a long polymer molecule. The equilibrium distribution of  Eq.(1) can be approximated by the continuous Gaussian distribution ($x = 2 b j$)
\begin{equation}
\label{de}
P_{0}(x)=\frac{e^{-\frac{x^2}{4 b^2 N}}}{(4 \pi b^2 N)^{1/2}}.
\end{equation}
Now if the individual monomers are further reduced to close points, the whole polymer can be represented in a continuum limit which results in Eq. (\ref{de}) as its equilibrium distribution. Then the corresponding probability conservation equation is given below
\begin{equation}
\tau_{R}\frac{\partial P(x,t)}{\partial t} = \left(4 N b^2 \frac{\partial^2}{\partial x^2} + 2 \frac{\partial}{\partial x} x \right) P(x,t).
\end{equation}
In the section this equation will be used to derive an analytical expression for end-to-end auto-correlation function.

\section{Analytical Calculation of end-to-end distance auto-correlation function}

\noindent Now we calculate the real time end-to-end distance auto-correlation function $\langle x(t)x(0)\rangle$ using Eq. (5). This auto-correlation function is defined by

\begin{equation}
\langle x(t) x(0)\rangle = \int_{-\infty}^{\infty} dx \; x \; \int_{-\infty}^{\infty} dx' \;x'\; G(x,x'|t)P_{0}(x'),
\end{equation}
where $P_{0} (x)$ is given by Eq.(4) and Green's function $G(x,x'|t)$ is to be determined form Eq.(5) as given below
\begin{equation}
\tau_{R} \frac{\partial}{\partial t} G(x,x'|t) = \left(4 N b^2 \frac{\partial^2}{\partial x^2}+2 \frac{\partial}{\partial x} x \right) G(x,x'|t),
\end{equation}
where the initial distribution is assumed to be given by
\begin{equation}
G(x,x'|0)= \delta(x-x').
\end{equation}
In the following, we will show that $\langle x(t)x(0)\rangle$ can be calculated without having the complete knowledge of $G(x,x'|t)$. Using Eq. (7) the time derivative of auto-correlation function in Eq. (6) is calculated as
\begin{equation}
\frac{\partial}{\partial t}\langle x(t)x(0)\rangle
=\int dx \int dx' \;x\;x'\;\left[\frac{1}{\tau_{R}}\frac{\partial}{\partial x}\left(4 N b^2\frac{\partial G(x,x'|t)}{\partial x} + 2 \;x \; G(x,x'|t) \right)\right] P_{0}(x').
\end{equation}
The right-hand side of the above equation can be re-written by the method of integration by parts as given by
\begin{eqnarray}
\frac{\partial}{\partial t}\langle x(t)x(0)\rangle = &
\int dx \int dx' G(x,x'|t) P_{0}(x')\left[\frac{4 N b^2}{\tau_{R}}\frac{\partial}{\partial x}\frac{\partial}{\partial x}(\;x\;x'\;)-\frac{\;2\;x}{\tau_{R}}\frac{\partial}{\partial x}(\;x\;x')\right] \\ \nonumber
& =-\frac{2}{\tau_{R}}\int dx \int dx' \;x\;x'\; G(x,x';t)P_{0}(x').  \nonumber
\end{eqnarray}
\noindent
The initial condition can be obtained from the following equation
\begin{equation}
\langle x(0)^2\rangle =\int dx \; x^2 \; P_{0}(x)= 2 N b^2.
\end{equation}
So we get the analytical expression for the end-to-end auto-correlation function s given by
\begin{equation}
\langle x(t)x(0)\rangle= 2 N b^2 e^{ - 2 t /{\tau_R}}.
\end{equation}

\section{Summary and Conclusions}

\noindent Here we use a very simple model to derive an analytical expression for end-to-end distance auto-correlation function for a long chain polymer molecule in solution. This auto-correlation decreases with increase in time and it is very sensitive to $\tau_{R}$, relaxation time from one configuration to another. Stronger correlation is found for polymer with longer bond lengths. Strong correlation is found for longer polymer chain.

\section{Acknowledgments:}
\noindent One of the author (M.G.) would like to thank IIT Mandi for HTRA fellowship and the other author thanks IIT mandi for providing PDA grant.


\section {References}
\begin{thebibliography}{99}
\bibitem{Szabo-Physica} K. Schulten, Z. Schulten and A. Szabo, Physica A 100 (1980) 599.
\bibitem{M-Physica} M. Ganguly and A. Chakraborty, Physica A, {484} (2017) {163}.
\bibitem{M-CPL} M. Ganguly and A. Chakraborty, Chem. Phys. Lett., (under revision) (2017).
\bibitem{Winnik} A. Winnik, In Cyclic Polymers, Chapter 9 Elsivier, 1986.
\bibitem{Haung} Z. Haung, H. Ji, J. Mays, and M. Dadmun, Langmuir 26 (2010) 202.
\bibitem{Lapidus} L. J. Lapidus, P. J. Steinbach, W. A. Eaton, A. Szabo, and J. Hofrichter, J. Phys. Chem. B 106 (2002) 11628.
\bibitem{Wilemski} G. Wilemski and M. Fixman, J. Chem. Phys. 60 (1974) 866.
\bibitem{Doi} M. Doi, Chem. Phys. 9 (1975) 455.
\bibitem{Szabo} A. Szabo, K. Schulten, and Z. Schulten, J. Chem. Phys. 72 (1980) 4350.
\bibitem{Pastor}R. W. Pastor, R. Zwanzig, and A. Szabo, J. Chem. Phys. 105 (1996) 3878.
\bibitem{Portman}J. J. Portman, J. Chem. Phys. 118 (2003) 2381.
\bibitem{Sokolov} I. M. Sokolov, Phys. Rev. Lett. 90 (2003) 080601.
\bibitem{Toan} N. M. Toan, G. Morrison, C. Hyeon, and D. Thirumalai, J. Phys. Chem. B 112 (2008) 6094.
\end{thebibliography}
\end{document}